# Reflexion in mathematical models of decision-making

This work is partially supported by the Russian Scientific Foundation, project no. 16-19-10609

Novikov Dmitry

*Director of V.A. Trapeznikov Institute of Control Sciences of Russian Academy of Sciences, Moscow, Russian Federation*

65 Profsoyuznaya street, Moscow 117997, Russia.

Korepanov Vsevolod

*Laboratory of Active Systems, V.A. Trapeznikov Institute of Control Sciences of Russian Academy of Sciences, Moscow, Russian Federation*

65 Profsoyuznaya street, Moscow 117997, Russia, vkorepanov@ipu.ru

Chkhartishvili Alexander

*Laboratory of Active Systems, V.A. Trapeznikov Institute of Control Sciences of Russian Academy of Sciences, Moscow, Russian Federation*

65 Profsoyuznaya street, Moscow 117997, Russia.

# Reflexion in mathematical models of decision-making

The paper is devoted to a survey of static and dynamic mathematical models of behavior with explicitly stated reflexive models of agents' decision-making. Reflexion is considered as agent's beliefs about nature, opponents' beliefs and opponents' decision-making principles in the framework of game theory, collective behavior theory and learning models.

This work is partially supported by the Russian Scientific Foundation, project no. 16-19-10609

## 1 About reflexion

One of the fundamental properties of human existence is that, along with the natural ('objective') reality, there exist its images in human minds. A purposeful study of this phenomenon is traditionally associated with the term 'reflexion'. The term reflexion (from Latin reflex 'bent back'; was first suggested by J. Locke) means [1]:

- a principle of human thinking, guiding humans towards comprehension and perception of one's own forms and premises;
- subjective consideration of a knowledge, critical analysis of its content and cognition methods;
- the activity of self-actualization, revealing the internal structure and specifics of spiritual world of a human.

Consider the situation with one subject. He has beliefs about the natural reality, but he can also realize these beliefs (construct beliefs about beliefs), etc. This is how the reflexive reality is formed. The subject's reflexion on his own beliefs about reality, the principles of his activity, etc. is called self-reflexion or a reflexion of the first kind. Reflexion of the second kind takes place in the situation of several subjects who know about the existence of each other and consists of beliefs about reality, the principles of decision-making, self-reflexion, etc. other subjects.

Since reflexion is a fundamental property of human, it is necessary to take it into account in the sciences studying the activity of a person and society, including the control sciences, in which human (people) acts as the object of control or a part of it. At the same time, we can assume that reflexion is present in such models in some aspects by default, if human (people) is modeled in them as an active element and / or control is present (see [2]). Methodological approaches to the correlation of reflexion and control in terms of the classical input-output control scheme are described in more detail in [2].

The aim of this paper is to review the existing approaches to the explicit consideration of reflexion in static and dynamic mathematical models of behavior from game theory, theory of collective behavior and behavioral economics, i.e., at first, an approaches, where reflexion is separate aspect of modeling, at second, an approaches, where reflexion can be easily identified by analogy with the first approaches.

The paper is organized as follows. Section 2 describes static reflexive models of noncooperative game theory. Section 3 provides dynamic models which includes reflexion. Some other examples presented in Section 4 and Section 5 concludes our work.

## 2 Statics. Noncooperative game theory in normal form

Game theory, branch of applied mathematics, that analyzes models of decision making in the conditions of noncoinciding interests of parties (subjects); each side strives for influencing the situation in his/her favor [3,4].

Further, to denote the subject making decision, and its decision, the terms agent and action are used. The main task of game theory is to describe the interaction of several agents whose interests do not coincide, and the results of activities (payofs, utility) of each depend in general on the actions of all. The result of such a description is the forecast of a reasonable and 'sustainable' outcome of the game - the so-called *game solution (equilibrium)*.

The basic description of the game is to set the following parameters:

- a set of agents $N = \{1, \dots, n\}$;
- a set of feasible actions of agents: $X_i, X = X_1 \times \dots X_n$. Further it will be understood that the sets of actions of agents are finite, unless the contrary is specified;
- preferences of agents (relationships between actions and payoffs). Each agent is supposed to strive for maximizing his/her payoff (the behavior of each agent appears purposeful): $u_i(x_1, \dots, x_n) \underset{x_i}{\to} max, x_i \in X_i$, $u_i$ is utility function of agent *i*;
- awareness of agents (information on essential parameters, being available to agents at the moment of their choice);

The above parameters define a game, but they are insufficient to predict its outcome, i.e., a solution (or an equilibrium). Nowadays, game theory suggests no universal concept of equilibria. By adopting different assumptions regarding principles of agent's decision making, one can construct different solutions. Thus, design an equilibrium concept forms a basic problem for any game-theoretic research.

In game theory, psychology, distributed systems and other fields of science (see the overviews in [5,6]), one should consider not only agents' beliefs about essential parameters, but also their beliefs about the beliefs of other agents, etc. The set of such beliefs is called the *hierarchy of beliefs*. We will model it using the tree of awareness structure of a reflexive game (see below). In other words, situations of interactive decision making (modeled in game theory) require that each agent 'forecasts' opponents' behavior prior to his/her choice. And so, each agent should possess definite beliefs about the view of the game by his/her opponents. On the other hand, opponents should do the same. Consequently, the uncertainty regarding the game to-be-played generates an infinite hierarchy of beliefs of game participants.

A special case of awareness concerns *common knowledge* when beliefs of all orders coincide (see [2] for a survey). In [7] was introduced a rigorous definition of common knowledge as a fact with the following properties:

(1) All agents know it;
(2) All agents know 1;
(3) All agents know 2 and so on – ad infinitum.

As surveyed in [2], there are two 'extremes' in the awareness of agents:

- The first 'extreme' is common knowledge;
- The second 'extreme' is the infinite hierarchy of compatible or incompatible beliefs given in [8]. On the one hand, it describes all possible Bayesian games and all possible hierarchies of beliefs. On the other hand, it appears very general and, consequently, very cumbersome, thus interfering with constructive statement and solution of specific problems.

Reflexive game is defined as a game in which agents make decisions based on the hierarchy of their beliefs. The dependence of game solutions on a finite hierarchy of compatible or incompatible beliefs of agents (the whole range between the above 'extremes') has been studied in [9,1].

According to game theory and reflexive models of decision making, it seems reasonable to distinguish between informational reflexion and strategic reflexion.

Informational reflexion is the process and result of agent's thinking about (a) the values of uncertain parameters and (b) what his/her opponents (other agents) know about these values. Here the 'game' component actually disappears–an agent makes no decisions. Strategic reflexion is the process and result of agent's thinking about which decision making principles his/her opponents employ under the awareness assigned by him/her via informational reflexion. Therefore, informational reflexion often relates to insufficient mutual awareness, and its result serves for strategic reflexion and then for decision making. Strategic reflexion takes place even in the case of complete awareness, precessing agent's choice of an action. It allows to eliminate the uncertainty about the behavior of opponents. In other words, informational and strategic reflexion can be studied independently, but the both occur in the case of incomplete or insufficient awareness.

Let's move on to a description of informational reflexion model.

## 2.1 Informational reflexion

Consider the set of agents: $N = \{1, 2, \ldots, n\}$. Denote by $\theta \in \Theta$ the uncertain parameter (let's the set $\Theta$ is a common knowledge for all agents) from which (in addition to actions) the agents utility function depends. The *awareness structure* $I_i$ of agent $i$ includes the following elements. First, the belief of agent $i$ about the parameter $\theta$; denote it by $\theta_i \in \Theta$. Second, the beliefs of agent $i$ about the beliefs of the other agents about the parameter ; denote them by $\theta_{ij} \in \Theta, j \in N$. Third, the beliefs of agent $i$ about the beliefs of agent $j$ about the belief of agent $k$; denote them by $\theta_{ijk} \in \Theta, j, k \in N$. And so on (evidently, this reasoning is generally infinite).

In the sequel, we employ the term 'awareness structure', which is a synonym of 'informational structure' and 'hierarchy of beliefs'. Therefore, the awareness structure $I_i$ of agent $i$ is specified by the set of values $\theta_{ij_1\ldots j_l} \in \Theta$, where $l$ runs over the set of nonnegative integer numbers, $j_1, \ldots, j_l \in N$.

The awareness structure $I$ of the whole game is defined in a similar manner; in particular, the set of the values $\theta_{i_1\ldots i_l} \in \Theta$, with $l$ running over the set of nonnegative integer numbers, $i_1, \ldots, i_l \in N$. We emphasize that the agents are not aware of the whole structure $I$; each of them knows only a substructure $I_i$.

Thus, an awareness structure is an infinite n-tree (each non-leaf node has n descendants); the corresponding nodes of the tree describe specific awareness of real and phantom (see below) agents.

A reflexive game $\Gamma_I$ is a game defined by the following tuple:

$$\Gamma_I = \{N, (X_i)_{i\in N}, u_i(\cdot)_{i\in N}, I\}$$

Therefore, a reflexive game generalizes the notion of a normal-form game (determined by the tuple $\{N, (X_i)_{i\in N}, u_i(\cdot)_{i\in N}\}$) to the case when agents' awareness is presented by an hierarchy of their beliefs (i.e., the awareness structure $I$). Within the framework of the accepted definition, a 'classical' normal-form game is a special case of a reflexive game (a game under a common knowledge among the agents). Consider the 'extreme' case when the state of nature appears a common knowledge; for a reflexive game, the solution concept (based on an informational equilibrium, see below) turns out equivalent to the Nash equilibrium concept.

To proceed and formulate a series of definitions and properties, we introduce the following notation:

- $\Sigma_+$ – stands for a set of finite sequences of indexes from $N$;
- $\Sigma$ is the $\Sigma_+$ with empty sequence appended;

- $|\sigma|$ – indicates the number of indexes in the sequence $\sigma \in \Sigma$ (for the empty sequence, it equals zero); this parameter is known as the length of an index sequence.

Imagine $\theta_i$ represents the belief of agent i about the uncertain parameter, while $\theta_{ii}$ means the belief of agent $i$ about his/her own belief. It seems then natural that $\theta_{ii} = \theta_i$. In other words, agent $i$ is well-informed on his/her own beliefs. Moreover, he/she assumes that the rest agents possess the same property. Formally, this means that the axiom of self-awareness is accepted: $\forall\, i \in N, \forall\, \tau, \sigma \in \Sigma: \theta_{\tau ii\sigma} = \theta_{\tau i\sigma}$.

In addition to the awareness structures $I_i (i \in N)$, one may also analyze the awareness structures $I_{ij}$ (i.e., the awareness of agent $j$ according to the belief of agent $i$), $I_{ijk}$, and so on. Let us identify the awareness structure with the agent being characterized by it. In this case, one may claim that $n$ real agents ($i$-agents, where $i \in N$) having the awareness structures $I_i$ also play with phantom agents ($\tau$-agents, where $\tau \in \Sigma_+, |\tau| \geq 2$) having the awareness structures $I_\tau = \{\theta_{\tau\sigma}\}, \sigma \in \Sigma$. It should be emphasized that phantom agents exist merely in the minds of real agents; still, they have an impact on their actions; these aspects will be explained below.

Assume that the awareness structure $I$ of a game is given; this means that the awareness structures are also defined for all (real and phantom) agents. Within the framework of the hypothesis of rational behavior, the choice of an action $x_\tau$ performed by a $\tau$-agent is described by his/her awareness structure $I_\tau$. Hence, the mentioned structure being available, one may model agent's reasoning and evaluate his/her action. On the other hand, while choosing his/her action, the agent models actions of the rest agents (i.e., performs reflexion). Therefore, estimating the game outcome, we should account for the actions of real and phantom agents.

A set of actions $x_\tau^*$, $\tau \in \Sigma_+$, is called an informational equilibrium, if the following conditions are met:

(1) The awareness structure $I$ possesses finite complexity [10];

(2) $\forall \lambda, \mu \in \Sigma_+\ I_\lambda = I_\mu \Rightarrow x_\lambda^* = x_\mu^*$;

(3) $\forall i \in N, \sigma \in \Sigma: x_{\sigma i}^* \in \underset{x_i \in X_i}{\mathrm{Argmax}}\, u_i(\theta_{\sigma i}, x_{\sigma i1}^*, \ldots, x_{\sigma i(i-1)}^*, x_i, x_{\sigma i(i+1)}^*, \ldots, x_{\sigma in}^*)$.

Here Condition 1 claims that a reflexive game involves a finite number of real and phantom agents (what happens when this assumption is rejected, is discussed in [11]). Condition 2 expresses the requirement that the agents with an identical awareness choose identical actions. Finally, Condition 3 expresses rational behavior of agents– each agent strives for maximizing the individual goal function via a proper choice of his/her action. For this, an agent substitutes actions of the opponents into his/her goal function; the actions are rational in the view of the considered agent (according to the available beliefs about the rest agents).

## 2.2 Strategic reflexion

Publications on strategic reflexion models (strategic thinking in English-speaking environment), the so-called level-k models, appeared in the mid-1990s [12,13,14]. The basic idea is that an agent can assume (when making a decision) that all other agents act uniformly randomly, then his best response will correspond to the maximization of the mathematical expectation of his utility function. Such an agent has rank 1 of reflexion, agents acting randomly – rank 0. Another agent may think that this type of beliefs will be for everyone, i.e. the other agents have 1 rank, and not 0, then its action maximizes his utility function, provided that the others act as agents of rank 1. Such an agent has rank 2. It is easy to construct an agent of arbitrary rank $k$.

In 2002, a generalization arose - a model of cognitive hierarchies (the Cognitive Hierarchies Model - CH) [15,16], in which the distribution of agents' ranks of reflexion is according to the Poisson distribution and an agent with rank $k$ thinks that other agents have not only rank $(k-1)$ but some 'truncated' or normalized Poisson distribution of $0, 1, \ldots, (k-1)$ ranks. Further in 2016, the authors constructed a generalized model of cognitive hierarchies [17], in which two free parameters were introduced into the model.

In parallel with this, the quantum best response (QBR) model developed [18], the idea of which is that people make mistakes the more often, the less these mistakes cost.

Stahl и Whilson [14] combined the level-k and QBR models: in the level-k model, players use QBR instead of the best response. In [19] the same idea was used, only CH was taken instead of the level-k model, which gave an effect - the QCH model more effectively describes the behavior of people in experiments. Further, work [200] shows that a small shift of the Poisson distribution underlying CH is also fruitful, the proposed Spike-Poisson QCH model (QCH with shifted Poisson distribution) has shown greater efficiency.

In fact, all these models have the four common components, variations of which cause their difference:

(1) Partitions of agents to the ranks of reflexion;
(2) Awareness of agent with rank $k$;
(3) The rank $k$ agent's response model to expected actions of opponents
(4) Behavior model of rank 0 agent.

**Partitions of agents to the ranks of reflexion model**. Let's define $\aleph = \{N^0, \ldots, N^m\}$ –

the set of agents $N$ partitions to reflexion ranks, where $N^i$ – the set of rank $i$ agents, $i = 0, \ldots, m$ – *maximum reflexion rank*. Let $n^i = |N^i|, \sum_0^m n^i = n$. Partition ℵ is called a *reflexive partition* [21]. We also introduce the notation $f^k = n^k/n$, then $f^0, \ldots, f^m$ can be called a *distribution of ranks*.

In the level-k model, it is usually assumed that $m = 2$, and the reflexive partitioning (three parameters $n^0, n^1, n^2$, of which two are free) are searched based on the criterion of the maximum descriptive power of the model on real data.

In the (generalized) model of Cognitive Hierarchies, $m$ is not limited or not greater than 7, and the distribution of ranks is the Poisson distribution with one parameter.

**Awareness of agent with rank k** is how he constructs his subjective reflexive partitioning of the set of agents. The general assumption among the models of strategic reflexion is that the agent of rank $k$ does not allow the existence of agents having the same or higher rank of reflexion (although the agent may think that there are agents of the same rank, but this is not always appropriate for various reasons, see Section 2.5 in [16]). Then his belief about reflexive partition will differ from the real ℵ.

For agent $j$ of rank $k$ the subjective reflexive partition is:

$$ℵ_j^k = \left(N^{k0}, N^{k1}, \ldots, N^{k(k-1)}, \{j\}, \emptyset, \ldots, \emptyset\right),$$

where $N^{kl}$ –set of agents of rank $l$ in beliefs of agent of rank $k$.

In the level-k model

$$N^{k(k-1)} = N\backslash\{j\}, N^{kp} = \emptyset, \forall p = 0, \ldots, k - 2$$

In the Cognitive Hierarchies model

$$f^{kp} = \frac{f^p}{\sum_{l=0}^{k-1} f^l}$$

And it is assumed that $N^p \subset N^{kp}$.

In the generalized cognitive hierarchy model

$$f^{kp} = \frac{(f^p)^\alpha}{\sum_{l=0}^{k-1}(f^l)^\alpha}, \alpha \geq 1$$

In the reflexive partitions method example from [22]:

$$N^{k(k-1)} = N^{k-1} \cup N^k\setminus\{j\} \cup N^{k+1} \cup \ldots \cup N^m, k \geq 1$$

$$N^{kp} = N^p, p < k-1, k \geq 2$$

**The rank $k$ agent's response model to expected actions of opponents** – with the given representations of opponents, the agent knows his action and can choose his action focusing on the utility function - choose the best answer or the 'quantum' best answer.

Let $s_j^k$ – a mixed strategy of agent $j$ of rank $k$. Let rank 0 agent choose a mixed strategy with equal probabilities for all actions $s_j^0 = s^U$, so that

$$\forall a_j \in X_j : s_j^0(a_j) = 1/|X_j|.$$

Definition. Let's define $u_j(a_i, s_{-j})$ – expected utiltity of $j$-th agent when playing action $a_i \in X_i$ against opponents' strategy profile $s_{-j}$. Then a *quantal best response* $QBR_i(s_{-i}; \lambda)$ by agent $i$ to $s_{-i}$ is a mixed strategy $s_i$ such that

$$s_i(a_i) = \frac{\exp[\lambda \cdot u_i(a_i, s_{-i})]}{\sum_{a_i'} \exp[\lambda \cdot u_i(a_i', s_{-i})]},$$

where $\lambda$ (the precision) represents agents' sensitivity to utility differences.

Most papers use such either best response or QBR model of response to expected actions:

$$r_j(s_{-j}) \stackrel{\text{def}}{=} \begin{cases} U(\underset{a \in X_i}{\text{Argmax}}(u_j(a, s_{-j}))), if\ best\ response\ model \\ QBR_j(s_{-j}, \lambda), \ if\ quantal\ best\ response\ model \end{cases},$$

where $U(A)$ – mixed strategy with equal probabilities for elements of $A \subseteq X$.

Then for agent of rank 1:

$$s_j^1 = r(s_{-j}^U).$$

For an agent of rank $k$:

$$s_j^k(\aleph_j^k) = r_j\left(\left(s_{N_{k0}}^0, \ldots, s_{N_{k(k-1)}}^{k-1}\right)\right)$$

**Behavior model of rank 0 agent**. In most works, it is assumed that an agent of rank 0 makes a uniform random choice between his actions. Such a model has the merit in that it is defined for any set of actions and sufficiently general. However, to describe the behavior of real people it is not always successful. For example, in [22] it was suggested to change the model of players of rank 0 that "dramatically improve predictions of human behavior, since a substantial fraction of agents may play rank-0 strategies directly, and furthermore since iterative models ground all higher-level strategies in responses to the rank-0 strategy" [22,p.1].

Variants of behavior of the 0 rank agent used in other works:

- maximin (pessimistic model), maximax (optimistic), minimax regret, fairness and efficiency in [22];
- averse to receiving minimum payoffs in [17];

- iterated dominance [23].

**The reflexive structure** is determined by a combination of the first two components of the approaches, and if an awareness model of an rank $k$ agent is given, then the reflexive structure is uniquely determined by the reflexive partition $\aleph$. Note that the reflexive structure considered within the framework of the models of strategic reflexion is in a sense similar to the information structure used in the models of information reflexion.

Agents' strategy profile

$$s^*(\aleph) = \left\{s_j^k\left(\aleph_j^k\right)\right\}_{j \in N_k, k=0..m}$$

are called reflexive equilibrium of the game $\Gamma_\aleph = \{N, u_i(\cdot)_{i \in N}, \aleph\}$ [21,24].

Without using the quantal best response, a reflexive equilibrium is usually possible to formulate in pure strategies. Reflexive equilibrium is the aggregate of actions of agents, which are the best answers to the actions of opponents assumed within bounds of the existing reflexive structure. Note that the reflexive equilibrium is quite exotic in the sense that in it the actions of agents in the general case are not the best responses to the actions chosen by opponents and it always exists by construction. A systematic classification of strategic reflexion models is given in [21, 24].

*2.3 Combined model of information and strategic reflexion*

Information and strategic reflexion models are similar, so the idea of combining them appeared and proposed in [25]. Such combined model is formulated for case of small number of agents, when beliefs of each can be clearly expressed.

Beliefs of an agent can be represented as an oriented tree whose nodes are agents, real or phantom (i.e. existing only in the mind of other agents). An arrow from agent A to agent B would mean that agent B knows all information about agent A (hereinafter, we will call a real or phantom agent just an 'agent', unless the opposite is stated).

Such a tree of agent 1, as an example, for $n = 3$, is shown on Fig. 1.

[Figure 1 near here]

On the diagram, node '1' is a real agent 1, nodes '12' ('one-two', not twelve) and '13' are agents 2 and 3 in beliefs of agent 1.

Note that nodes '12' and '13' are 'hanging', i.e. that phantom agents 12 and 13 do not have beliefs about opponents. This fact means that such agents have rank 0. As stated above the rank of an agent is one greater than the largest rank that his opponents have, in his opinion. So agent 1 on Fig. 1 has rank 1.

If agent 1 thinks that his opponents have rank 1, for example, then he has rank 2. Similarly, if agent 1 thinks that his opponents 2 and 3 have ranks 0 and 1, then he also has rank 2. Graphs for these cases are shown on Fig. 2(a) and 2(b).

[Figure 2(a) and 2(b) near here]

Thus, the rank of an agent equals to the number of arrows in maximal length path from leaf nodes to the agent (equals 0 if he is a leaf node of a tree - there are no incoming arrows). So agents of rank 2 and more must have beliefs about beliefs of their opponents and so on (these are agents 121, 123 etc. on Fig. 2).

For a complete description of a situation we need to define beliefs of all agents which creates a hierarchic information structure. An example of such a structure is given in Fig. 3.

[Figure 3 near here]

We can merge some agents as they have equal beliefs and, consequently, equal actions.

For example, 1212 and 132 on Fig. 3 are phantom agents 2 with rank 0 who act identically, so we can merge them in one node. In other words, agents 121 and 13 have equal beliefs about agent 2. Furthermore, if we identify in this way all rank 0 agents, we can see that agents 12 and 2 have equal beliefs about agent 3 with rank 1.

After completing all merging operations, we get a more visually attractive graph of agents' beliefs, *the graph of a reflexive game* [1], where each node represents a real or a phantom agent (Fig. 4).

[Figure 4 near here]

As we can see, the graph of a reflexive game displays the following information:

- ranks of agents;
- adequate or inadeqate beliefs: agent 1 is adequately informed about agent 3 (i.e. the image of agent 3 in mind of agent 1 matches with real agent 3), but he is not adequately informed about the agent 2.
- equivalent beliefs: agents 3 and 12 have the same beliefs about the agent 1 (though incorrect).

Further, the information reflexion to the existing hierarchic beliefs structure is added. Let $\theta \in \Theta$ be the external system parameter on which agents' utility functions are dependent on: $u_i: \Theta \times X \to \mathbb{R}$. The second assumption is that agents' have common knowledge that the parameter value is not known exactly – each agent can estimate it differently. In other words agent $i$ have belief about $\theta$: $\theta_i \in \Theta$.

Now, to predict agent $j$ action an agent $i$ must have an opinion about opponent's beliefs about the value of $\theta$: $\theta_{ij} \in \Theta$ and his beliefs about his opponents beliefs and so on. We can put simple information reflexion model to our example case (Fig. 4). Let all agents believe that $\theta = a \in \Theta$. Then it can be displayed as the same graph as in Fig. 4 – equal beliefs of all are not shown by convention. Here $\theta_1 = \theta_2 = \theta_3 = \theta_{12} = \theta_{23} = \theta_{31} = \theta_{21} = \theta_{32} = \theta_{313} = a$.

Let all agents believe that $\theta = a$ but agent 23 believes that $\theta = b \neq a$. Now we cannot merge agents 23 and 123 since they have different beliefs (equal strategic but different information reflexion), and so they will be represented by different nodes in the graph of a reflexive game (opposite to the situation in Fig. 4). This is shown on Fig. 5 (here, beliefs of most agents $\theta = a$ are not shown). Now agents 2 and 12 have different beliefs about agent 3. Despite the fact that agent 23 thinks that $\theta = b$, he believes that his opponents (agents 21 and 32) think that $\theta = a$.

[Figure 5 near here]

It is clear that this way we can specify an arbitrary beliefs structure about the parameter $\theta$, i.e. such a structure in which all agents have definite beliefs about $\theta$ and about the opponents' beliefs about $\theta$ and so on, see [1]. The resulting model is a combination of the information reflexion model and the strategic reflexion model.

**3 Dynamics. Collective behavior models, learning and teaching models**

The simplest model in moving from a static situation to a dynamic one is a *repeated game* of $n$ agents, in which:

(1) A fixed number of decision-making stages $t \in \{0,1, \dots, T\}$,

(2) At each stage, each agent $i$ chooses his action $x_i(t) \in X_i$,

(3) Each stage has its own result – payoff of each agent $u_i(x_1(t), \dots, x_n(t)) \in \mathbb{R}$,

(4) An agent's payoff in each stage does not depend on the stage (time) and the history (sequence of actions), but only on the current stage actions: $u_i: X \to \mathbb{R}$,

(5) An agent's overall payoff is a total payoff in each stage.

Static models can be considered as a starting point for dynamics of agents - such models that allow you to select the action in the first stage of a repeated game. In spite of condition 4, a rational player can take into account history of actions to identify opponents behavior.

In this chapter, we first consider the collective behavior approach, in which the behavior of a group of agents is studied, built on the local boundedly rational behavior of agents of the group, at each stage tending to maximize utility functions. Then learning models, where each agent is trying to find the optimal mixed strategy by learning from received payoffs and/or game history. Then we will mention models with reflexion – teaching models - if the agent knows that the others are learned, then he can use it to teach them profitable for himself strategies.

*3.1 Collective behavior models*

Traditionally, in game-theoretic models and/or in models of collective decision-making, one of two assumptions about mutual awareness of agents is used [24]. Either it is believed that all the essential information and the principles of decision-making by agents are known to all, everyone knows that everyone knows this, etc. to infinity. In other words the concept of common knowledge is used. Either it is assumed that each agent, within the bounds of his awareness, follows a certain procedure for making individual decisions and almost 'does not think' about what they know and how the other agents behave. The first approach is canonical for game theory, the second - for models of collective behavior, and also partly for learning models. But between these two 'extremes' there is a rather large variety of possible situations. Suppose that an agent in situation of common knowledge of significant external parameters (no information reflexion) carried out an act of strategic reflexion - he tried to predict the behavior (not only awareness, but also the principles of decision making) of other agents and chooses his actions with this forecast (such an agent has the first rank of reflexion). Another agent (having a second rank of reflexion) may know about the existence of agents of the first rank and predict their behavior. And so on.

Classic game-theoretic models proceed from the following. In a normal form game, agents choose Nash equilibrium actions. However, investigations in the field of experimental economics indicate this not always the case (e.g., see [26] and the overview [27]). The divergence between actual behavior and theoretical expectations has several explanations:

- limited cognitive capabilities of agents [28] (evaluation of a Nash equilibrium represents a cumbersome computational problem [29], especially decentralized). It should also be emphasized that the Nash equilibrium does not always adequately describe the real behavior of agents in laboratory experimental one-step games, including because agents do not have time to 'correct' their misconceptions about the essential parameters of game. For example, the concept of D. Bernheim's rationalized strategies requires unlimited rationality of agents [30];
- agent's full confidence in that all the opponents would evaluate a Nash equilibrium;
- incomplete awareness;
- the presence of several equilibria.

Therefore, there exist at least two foundations ('theoretical' and 'experimental' ones) for considering models of collective behavior of agents with different reflexion ranks.

In contrast to game theory, the theory of collective behavior analyzes the behavior dynamics of rational agents under rather weak assumptions regarding their awareness. For instance, far from always agents need a common knowledge about the set of agents, sets of feasible actions and goal functions of opponents. Alternatively, agents may not predict the behavior of their opponents (as in game theory). Moreover, making decisions, agents may know nothing about the existence of some agents or possess aggregated information about them.

The most widespread model of collective behavior dynamics is *the model of indicator behavior* (see references in [24]). The essence of the model consists in the following. Suppose that at stage *t* each agent observes the actions of all agents $\{x_i(t-1)\}_{i \in N}$, that have been chosen at the previous stage $t-1, t = 1,2, ...$ The initial action vector $x(0) = (x_1(0), ..., x_n(0))$ is assumed known.

Each agent can evaluate his/her *current goal* – an action maximizing his/her utility function provided that at a current stage all agents will not change actions from the previous stage (either action giving maximum is unique or we have a formal procedure to choose one of that):

$$w_i(x_{-i}(t-1)) = \underset{y \in X_i}{\operatorname{argmax}} u_i(y, x_{-i}(t-1)), i \in N, t = 1,2, .... \quad (1)$$

According to the hypothesis of indicator behavior, at each game stage an agent makes a 'step' from his/her previous action to the current goal:

$$x_i(t) = x_i(t-1) + \gamma_i^t(w_i(x_{-i}(t-1) + x_i(t-1)), i \in N, t = 1,2, ..., \quad (2)$$

where $\gamma_i^t \in [0,1]$ designate 'step size'.

It can be seen that expression (2) implies that the sets of actions of agents are vector spaces. This emphasizes the fact that models of collective behavior usually consider the movements of agents on a plane or in space. On the other hand, if the sets of actions are finite, the hypothesis of indicator behavior can be reformulated where mixed strategies of players are used as actions. Such a model will be to some learning models (see below).

For convenience, such collective behavior can be conditionally called 'optimization behavior', thereby emphasizing its difference from game behavior. The approaches of the theory of collective behavior and game theory are consistent in the sense that both investigate the behavior of rational agents, and game equilibria, as a rule, are equilibria of the dynamic procedures of collective behavior (for example, the Nash equilibrium is an equilibrium of the dynamics (2) of collective behavior).

To make the picture complete, note one more aspect, as well. The theory of collective behavior proposes another approach (going beyond the scope of this work), namely, evolutionary game theory [31]. This science studies the behavior of large homogeneous groups (populations) of individuals in typical repeated conflicts; each strategy is applied by a set of players, whereas a corresponding utility function characterizes the success of specific strategies (instead of specific participants of such interaction).

Thus, game theory often employs maximal assumptions regarding agents' awareness (e.g., the common knowledge hypothesis), while the theory of collective behavior involves the minimal assumptions. The addition of reflexion allows us to create an intermediate model, as was done in [22] using strategic reflexion, namely: within the hypothesis of indicator behavior (2), it is implicitly assumed that an agent, when choosing his actions, does not think about the fact that other agents act same. If he thought about it (carried out the reflexion), then at the time *t* he should have sought the best response to the actions of other agents predicted by expression (2). Thus, the position of the goal would no longer be determined by the expression (1), but by the expression $w_i(x_{-i}(t))$, where $x_{-i}(t)$ is determined by the expression (2).

In this case, it can be assumed that the reflexing agent of the first rank considers all others as non-reflexive. Analogously, we can consider agents of higher reflexion ranks too. For their description we use the reflexive partitions of a set of agents $N$: $\aleph = \{N^0, \ldots, N^m\}$ (see the strategic reflexion section above for formal description).

We will assume further that the agent with reflexion rank $k$ reliably knows the sets of agents of all lower ranks $k'$ (where $k' < k - 1$) and considers all agents of his and higher ranks ($k'' \geq k$) having rank one less than itself (that is, rank $k - 1$).

Let the vector $x(0)$ of the 'initial' actions of agents be given. Consider the following dynamic model of the reflexive decision making.

<u>Rank 0 of reflexion.</u> We will assume that agents with zero rank of reflexion (belonging to the set $N^0$) assume that the actions of the remaining agents will be the same as in the previous time:

$$x_i(t) = x_i(t-1) + \gamma_i^t[w_i(x_{-i}^{t-1}) - x_i^{t-1}], i \in N^0, t = 1,2,\ldots \quad (3)$$

<u>Rank 1 of reflexion.</u> Agent $j$, who has the first rank of reflexion ($j \in N^1$), considers all other agents to have zero-rank reflexion and, in accordance with (3), 'predicts' their choice. Therefore, his own choice $x_j^1(t)$ will be focused on the best response to the situation, which from his point of view should arise:

$$x_j^1(t) = x_j^1(t-1) + \gamma_j^t[w_i(x_{-i}(t)) - x_j^1(t-1)], j \in N^1$$

For the agent $j \in N^1$, the forecasted trajectory is defined by $(x(0), \ldots, (x_j^1(t), x_{-j}(t)), \ldots)$. The realized trajectory may not coincide with the trajectories predicted by agents of various degrees of reflexion [1].

<u>Rank 2 of reflexion.</u> We will assume that each agent $j$ having the second rank of reflexion ($j \in N^2$) knows the set $N^0$ reliably and considers all agents from the set

$N \setminus N^0 \setminus \{j\}$ possessing the first rank of reflexion. Because of this, he can 'predict' the behavior of all his opponents and his choice will be:

$$x_j^2(t) = x_j^2(t-1) + \gamma_j^t \left[ w_i \left( x_{N^0}(t), x_{N^1 \cup N^2 \{j\}}^1(t) \right) - x_j^2(t-1) \right], j \in N^2$$

For agent $j \in N^2$ the forecasted trajectory is defined by

$$(x(0), \ldots, (x_j^2(t), x_{N^0}(t), x_{N^1 \cup N^2 \setminus \{j\}}^1(t)), \ldots)$$

<u>Rank k of reflexion</u> ($k \leq m$). The behavior of rank $k$ agents is described similarly to the three cases (zero, first and second ranks of reflexion) considered above, taking into account the following reflexive structure of agents. Let $\aleph_j^k$ be the subjective reflexive partition – beliefs of rank $k$ agent $j$ about reflexive partitions of all agents:

$$\aleph_j^k = (\underbrace{N^0, N^1, \ldots, N^{k-2}, N^{k-1} \cup N^k \cup \ldots \cup N^m \setminus \{j\}}_{k}, \{j\}, \underbrace{\emptyset, \ldots, \emptyset}_{m-k-1}), j \in N^k$$

Rank $k$ agent $j$ will choose an action in accordance with expression

$$x_j^k(t) = x_j^k(t-1) + \gamma_j^t \left[ w_i(x_{N^0}(t), x_{N^1}^1(t), \ldots, x_{N^{k-1} \cup \ldots \cup N^m \setminus \{j\}}^{k-1}(t)) - x_j^k(t-1) \right]$$

## *3.2 Learning and teaching models*

In the basic formulation of the learning model, they rely on the idea of learning with reinforcement: the agent selects the action at each stage of the game and receives the result - a gain. The idea is that the agent should strive to choose those actions that led to big gains. Wherein game setting is absent, there is only an agent and an 'environment' (static or dynamically changing), acting in which the agent receives his gains. Nevertheless, experimental game theory successfully uses learning models to describe the behavior of players in games. Such models are called reinforcement-based models.

Such models 'grown up' in the field of individual decision making [32] and psychology, see [33].

In parallel, there is a 'game' formulation of learning models - players are not looking at their gains, but on the actions chosen by opponents (of course, if it is observable information case). That is, at each stage, using the history of actions of opponents, the agent builds, in general, a simple (statistical) model of the behavior of opponents and chooses an action that maximizes his utility function for a given behavior of opponents. Such models are called belief-based models, they have developed in the theory of games.

One can say that the first formulation does not use reflexion, and the latter explicitly introduces the agent's reflexion into the model - the decision-making model of opponents. It seems that these two models have nothing in common - they are based entirely on different principles and different information, but work of Camerer and Ho [34] showed that they have much in common. They proposed a model of learning, particular cases of which are models of one and the other side (also a short review on this topic was made).

Historically, one of the first belief-based models was the Cournot model [35], in which it was assumed that by choosing his action at the next stage, the player believes that opponents will not change their actions and choose the best response to the actions of opponents in the previous stage (like an indicator behavior with move size equal to 1). That is a simple model of response to a dynamically changing situation.

The next widely known model is the fictitious play model [36,37] (Fictitious Play, FP). Here the player forms a model of the behavior of opponents as a random variable with an empirical distribution function, observing their actions in the previous stages. Then the best response is chosen - the action that brings the maximum expected gain for the given behavior of the opponent. This model is based on the idea of learning, although it remains some 'static' in the understanding of opponents - using the empirical distribution function implies that the opponent does not change his behavior.

In addition to the above models, models in [38,39,40] also belong to this class, which are various generalizations of the Cournot and FP to overcome their weaknesses.

In the reinforcement-based models, the player does not need to know any information about the other players. Under sufficient assumptions, knowing only his gains from the game history allows them to correct next-stage action and find the optimal strategy, at least in asymptotic by stages. Models of this class presented in [33,41].

In [34] a hybrid EWA model (experience weighted attractions) with six free parameters was constructed. The values of the parameters under which the model is identical to reinforcement- and belief-based models are shown. A discussion of the qualitative value of the free parameters of the model is proposed. In [42] fixed values for three model parameters and functional dependencies that determine the values of two free parameters from the game's history are proposed. Thus, the number of free parameters is reduced to one as a response to criticism of a large number of free variables of the EWA model. It is shown, however, that such a model can well describe existing data of human behavior.

Finally, one of the recent models - the Individual Evolutionary Learning (IEL) model [43] - is proposed for a case of continuum cardinality set of actions. In it, the choice is also made from discrete set of actions, at each stage the elements of it can be replaced with a given probability by an element from the initial continuum set of player actions. The accumulation of the value of "reinforcements" occurs in a different way:

selecting two arbitrary actions, duplicating a more successful action, and removing a less successful one. That is why the model has in its name the word 'evolutionary'.

In the Sophistication model [44,45] reflexion was added in a manner similar to reflexion in the indicator behavior model (see above), namely: the reflexive agent believes that opponents use the learning model for decision making. Based on this model, he predicts their behavior at the next stage and chooses best response action. This is an agent of rank 1 reflexion. There is also a feature of the proposed model - players of the same rank know about each other and must choose the Nash or quantal response equilibrium. Agents of higher ranks are built in the same way as models of strategic reflexion.

In the same works ([44,45]) the teaching model are proposed. The agent believes that his opponents use the learning models and uses it not only to predict their actions, but also to choose a farsighted strategy of teaching opponents.

**4 Examples**

*4.1 Dynamic of information awareness*

A model describing the dynamics of agent awareness was considered in [46] on the basis of the following situation: three friends play a game in which the third pick two integers (possibly coincident) between 1 and 9 inclusive and reports sum of these numbers to the first, and their product to the second. Then the third one asks: 'what numbers are picked?'. The first and second should call these numbers or answer 'I do not know', they respond simultaneously and do not exchange any information, but they know history of answers. Both answered the same question: 'I do not know'. The third repeated the question: 'what numbers are picked'? The first and second, having thought, again answered: 'I do not know'. The third repeated the question and received the same answer. So it was repeated seven times, and on the eighth the first player named the numbers.

It turns out (see [46]) that a complete description of the awareness of the first and second players (including, in addition to the two real ones, 25 phantom agents) allows one to determine which numbers were picked and how the awareness changed after each answer.

*4.2 The diffuse bomb problem*

In work [47] so-called 'the diffuse bomb problem' are considered:

- The goal of a group consisting of several moving objects (agents) moving on a plane is a search (defeat) of a fixed target;
- The time to reach the goal is not fixed;
- There are several fixed sensors;
- When a sensor detects an agent, there is a risk of agent destruction increasing with decreasing distance between them;
- Agents move with a constant absolute value of a given speed, the direction of motion can vary;
- Planning of agent trajectories is carried out decentralized (autonomously) in real time.

The effeciency criterion of the actions of the group is their number $K$ that have reached the goal. Reflexive models are introduced in the formulation, where the risk of detection also increase with decreasing distance among agents, hence the behavior forecasts of the opponents is important.

The first reflexive behavior model is introduced similarly to the reflexion in the indicator model: reflex agents predict the action of the other agents in the next stage and take this into account when calculating the detection risk. In accordance with the introduced notation, this is a model of strategic reflexion, and only agents of rank 0 (non-reflexive) and 1 are introduced. Simulation experiments have shown that the efficiency of a group with proportion of rank 1 agents is increased, though not much. While the survival of reflexive agents (rank 1) is also higher than non-reflexive (rank 0).

The second reflexive behavior model is introduced in a situation where some agents do not know the level of risk. Those agents who know the level of risk, let's call them 'scouts', go to the goal first. Agents who do not know the level of risk (adaptive) observe the movement of the scouts and, knowing their behavior principles, build an empirical level of risk that they use for their own movement. Simulation experiments have shown that in this particular case, 20% of scouts (agents with large capabilities) are sufficient to ensure that the effectiveness of the group does not fall much compared to the situation when all agents are scouts. This model can be referred to as information reflexion, since the agents' beliefs about nature and their changes in the development of the situation are modeled.

In [48] the investigation of the diffuse bomb problem continues, but a game-theoretic component is introduced - two players simultaneously and independently choose their actions: one player selects the type of behavior model of agents, the second selects the sensors location. It is a matrix game. The Nash equilibrium in the original game and in the game of ranks, where players choose their ranks of strategic reflexion are considered. In a concrete example, it turned out that unique Nash equilibrium (in mixed strategies) does not change in the transition to a game of ranks. But in the game of ranks it is easier to find it because dimension of the game is less than the original game (not more in general case).

# 5 Conclusion

This review on the use of reflexion in static and dynamic formulations shows, on the one hand, the existence of common features of these approaches, on the other hand, the different possibilities of using reflexion in models:

- to analyze, predict and control game theoretic situation;
- to construct collective behavior;
- to learn and to teach opponents behavior.

Figure 1. An example of a belief tree.

Figure 2. (a) beliefs of agent 1 that his opponents are rank 1 agents, (b) beliefs of agent 1 that his opponents has rank 0 and 1.

Figure 3. All agents' beliefs example.

Figure 4. The graph of a reflexive game.

Figure 5. The graph of a reflexive game with information reflexion.